\documentclass[10pt]{article}
\usepackage{hyperref}
\hypersetup{colorlinks=true,linkcolor=black,citecolor=black,urlcolor=black,breaklinks=true}
\usepackage{float}
\usepackage[symbol]{footmisc}
\usepackage[superscript,sort]{cite}
\usepackage{array}
\usepackage{bm}
\usepackage{longtable}

\newcolumntype{x}[1]{%
	>{\centering\hspace{0pt}}p{#1}}%
\usepackage[normalem]{ulem} 
\usepackage{amsmath,amssymb}

\usepackage{amsthm,amscd,amsxtra,amsfonts,amsmath,amssymb,multirow}
\usepackage{wrapfig}
\usepackage[tiny,compact]{titlesec}
\usepackage{threeparttable} 
\usepackage{helvet}

\usepackage{booktabs}
\usepackage[table]{xcolor}
\usepackage{xcolor,colortbl}
\usepackage{placeins}
\usepackage{tikz}
\usetikzlibrary{shapes,arrows}
\usepackage[T1]{fontenc}
\usepackage[linesnumbered,ruled]{algorithm2e} 
\titlespacing*{\section}{0pt}{*0}{*0}
\titlespacing*{\subsection}{0pt}{*0}{*0}
\titlespacing*{\subsubsection}{0pt}{*0}{*0}
\titlespacing{\paragraph}{0pt}{*0}{*1}

\definecolor{MyPurple}{rgb}{1,0,1}

\newcommand{\beq}[1]{\begin{equation} \label{#1}}
\newcommand{\eeq}{\end{equation}}
\newcommand{\barray}{\begin{array}{ll}}
	\newcommand{\earray}{\end{array}}

\usepackage{longtable} 
\usepackage{threeparttable} 

\usepackage{array}
\newcolumntype{P}[1]{>{\centering\arraybackslash}p{#1}}

\usepackage{multirow}
\usepackage{color, colortbl}

\definecolor{Lightblue}{rgb}{0.867,0.914,0.961}
\definecolor{Lightgreen}{rgb}{0.883,0.934,0.848}

\DeclareGraphicsExtensions{.pdf,.jpeg,.png}

\title{Generative network complex (GNC) for drug discovery}
\author{Christopher Grow$^{1}$, Kaifu Gao$^{1}$, Duc Duy Nguyen$^{1}$,  and  Guo-Wei Wei$^{1,2,3,}$\footnote{
		Corresponding to Guo-Wei Wei.		Email: wei@math.msu.edu}\\
$^1$ Department of Mathematics,
Michigan State University, MI 48824, USA.\\
$^2$ Department of Electrical and Computer Engineering,
Michigan State University, MI 48824, USA. \\
$^3$ Department of Biochemistry and Molecular Biology,
Michigan State University, MI 48824, USA. \\
}

\date{\today}

\begin{document}

\maketitle

\begin{abstract}
It remains a challenging task to generate a vast variety of novel compounds with desirable pharmacological properties. In this work, a generative network complex (GNC) is proposed as a new platform for designing novel compounds, predicting their physical and chemical properties, and selecting potential drug candidates that fulfill various druggable criteria such as binding affinity, solubility, partition coefficient, etc. We combine a SMILES string generator, which consists of an encoder, a drug-property controlled or  regulated latent space, and a  decoder, with verification deep neural networks,  a target-specific three-dimensional (3D) pose generator, and mathematical deep learning networks to generate new compounds, predict their drug properties, construct 3D poses associated with target proteins, and reevaluate druggability, respectively.
{New compounds were generated in the latent space by either randomized  output,  controlled output, or optimized output.} In our demonstration, 2.08 million and 2.8 million novel compounds are generated respectively for Cathepsin S and BACE targets. These new compounds are very different from the seeds and cover a larger chemical space. For potentially active compounds,  their 3D poses are generated using a state-of-the-art method. The resulting 3D complexes are further evaluated for druggability by a championing deep learning algorithm based on algebraic topology, differential geometry, and algebraic graph theories. Performed on supercomputers, the whole process took less than one week. Therefore, our  GNC is an efficient new paradigm for discovering new drug candidates.

\end{abstract}

\section{Introduction}
Drug design and discovery ultimately test our understanding of biological sciences, the status of biotechnology,  and the maturity of  computational sciences and mathematics. Technically, drug discovery involves target discovery, lead discovery, lead optimization, preclinical development, three phases of clinical trials, and finally, launching to market only if everything goes well. Among them,  lead discovery, lead optimization, and preclinical development disqualify tens of thousands of molecules based on their binding affinities, solubilities, partition coefficients, clearances, permeabilities, toxicities, pharmacokinetics, etc., leaving only about ten compounds for clinical trails. Currently, drug discovery is both expensive and time-consuming. It takes about \$2.6 billion dollars and more than ten years, on average, to bring a new drug to the market. \cite{dimasi2016innovation}  Reducing the cost and speeding up the drug discovery process are crucial issues for the pharmaceutical industry. Much effort has been taken to optimize key steps of the drug discovery pipeline. For example, the development of high-throughput screening (HTS) has led to an unprecedented increase in the number of potential targets and leads \cite{hughes2011principles}. HTS is able to quickly conduct millions of tests to rapidly identify active compounds of interest using compound libraries \cite{macarron2011impact}.

\paragraph{}
While there has been an increase in the number of potential targets and leads, the number of new molecular entities generated has remained stable because of a high attrition rate during preclinical development and clinical phases, caused by the selection of leads with inappropriate physicochemical or pharmacological properties \cite{graul2009overcoming,tareq2010predictions}. Rational drug design (RDD) approaches are proposed to better identify candidates with the highest probability of success \cite{waring2015analysis}.  RDD aims at finding new medications based on the knowledge of biologically druggable targets \cite{dimasi2016innovation,stromgaard2017textbook}.
Several empirical metrics, such as Lipinski's rule of five (RO5) \cite{lipinski1997experimental}, were established for estimating druglikeness, which describes the druggability of a substance with respect to factors like bioavailability, solubility, toxicity, etc. Generally, the early selection of candidates requires the design of molecules complementary in shape and charge to the target of interest, which leads to a high binding affinity. Additionally, the determination of  the nature and  rates of physical/chemical/biological processes that are involved in the absorption, distribution, metabolism, and elimination (ADME) of drug candidates are also of primary importance. ADME profiling and prediction are mostly dependent on molecular descriptors such as RO5 \cite{tsaioun2009addme}. Furthermore, cellular/animal disease models are typically used during lead optimization to measure various pharmacokinetics. Finally, toxicity study is a primary task for preclinical development.

\paragraph{}
Recently, computer-aided drug design (CADD) has emerged as a useful approach in reducing the cost and   period of drug discovery \cite{alqahtani2017silico}. Computational techniques have been developed for both virtual screening (VS) and  optimizing the ADME properties of lead compounds. Essentially, these methods are designed as in silico filters to eliminate compounds with undesirable properties. These filters are widely applied for the assembly of compound libraries using combinatorial chemistry \cite{balakin2009compound}. The integration of early ADME profiling of lead chemicals has contributed to the speed-up of lead selection for phase-I trials without large amounts of revenue loss \cite{kapetanovic2008computer}. Currently, compounds are added in libraries on the basis of target-focused design or diversity considerations \cite{huang2016protein}. VS and HTS can screen compound libraries to select a subset of compounds whose properties are in agreement with various criteria \cite{szymanski2012adaptation}.

\paragraph{}
Despite these efforts, the current size of databases of chemical compounds remains small when compared with the chemical space spanned by all possible energetically stable stoichiometric combinations of atoms and topologies in molecules. Considering these factors, it is estimated that there are 10$^{60}$ distinct molecules. Among them, 10$^{30}$ are druglike \cite{macarron2011impact}. As a result, computational techniques are also being developed for the de novo design of druglike molecules \cite{schneider2005computer} and for generating large virtual chemical libraries, which can be more efficiently screened for in silico drug discovery.

\paragraph{}
Among the computational techniques available, deep neural networks (DNN) have gained much interest for their ability to extract  features and learn physical principles from training data. Currently, DNN-based architectures have been successfully developed for applications in a wide variety of fields in the biological and biomedical sciences \cite{min2017deep,mamoshina2016applications}.

\paragraph{}
More interestingly, several deep generative models based on variational autoencoders (VAEs) \cite{kingma2013auto}, adversarial autoencoders (AAEs) \cite{makhzani2015adversarial}, recurrent neural networks (RNNs) \cite{mandic2001recurrent},   long short term memory networks (LSTMs) \cite{hochreiter1997long} and generative adversarial networks (GANs) \cite{creswell2018generative} have been proposed for exploring the vast druglike chemical space. A policy-based reinforcement learning approach  was proposed to tune RNNs for episodic tasks \cite{olivecrona2017molecular,popova2018deep}. A VAE was used by Gomez-Bombarelli et al. \cite{gomez2018automatic} to encode a molecule in the continuous latent space for exploring associated properties. The usage of these models has been extended to generate molecules with desired properties  \cite{kang2018conditional}. Miha Skalic et al. \cite{skalic2019shape} combined a conditional variational autoencoder and a captioning network to generate previously unseen compounds from input voxelized molecular representations. Artur Kadurin et al. \cite{kadurin2017drugan} built an AAE to generate new compounds. Boris Sattarov et al. \cite{sattarov2019novo} combined deep autoencoder RNNs with generative topographic mapping to carry out de novo molecular design.

\paragraph{}
It is particularly interesting and important to generate potential drug candidates for specific drug targets.  To this end, a network complex is required to fulfill various functions, including target-specific molecular generation, target-specific binding affinity ranking, and solubility and partition coefficient evaluation. In this work, we propose a generative network complex (GNC) to combine drug-property controlled or regulated autoencoder (AE) models and DNN predictors to generate millions of new molecules and select potential drug candidates that have appropriate druggable properties. Our GNC includes the following components:
\begin{enumerate}
    \item Using known molecules in a target-specific training set as seeds, a SMILES string generator is constructed to generate millions of novel compounds. This generator consists of a CNN-based encoder, a drug-property controlled or regulated latent space, and a LSTM-based decoder.
    \item A pre-trained multitask DNN model is constructed to select  drug candidates based on druggable properties.
    \item A 3D structure generator, MathPose, to convert selected 2D SMILES strings into 3D structures based on target receipt information.
		\item A 3D multitask druggable property predictor, {mathematical deep learning (MathDL),} to further select new drug candidates via various druggable criteria.
\end{enumerate}
Some of these components, namely MathPose and MathDL, have been extensively validated in blind settings \cite{nguyen2019mathematical,nguyen2019D3R}. Our GNC can not only generate new molecules, but also   construct or pick up the molecules with ideal drug properties. This makes it a very promising method for generating millions of new drug candidates in silico in a very short time period.

\section{Methods }

\subsection{The structure of generative network complex (GNC)}
\begin{figure}[h]
    \centering
	\includegraphics[width=0.65\textwidth]{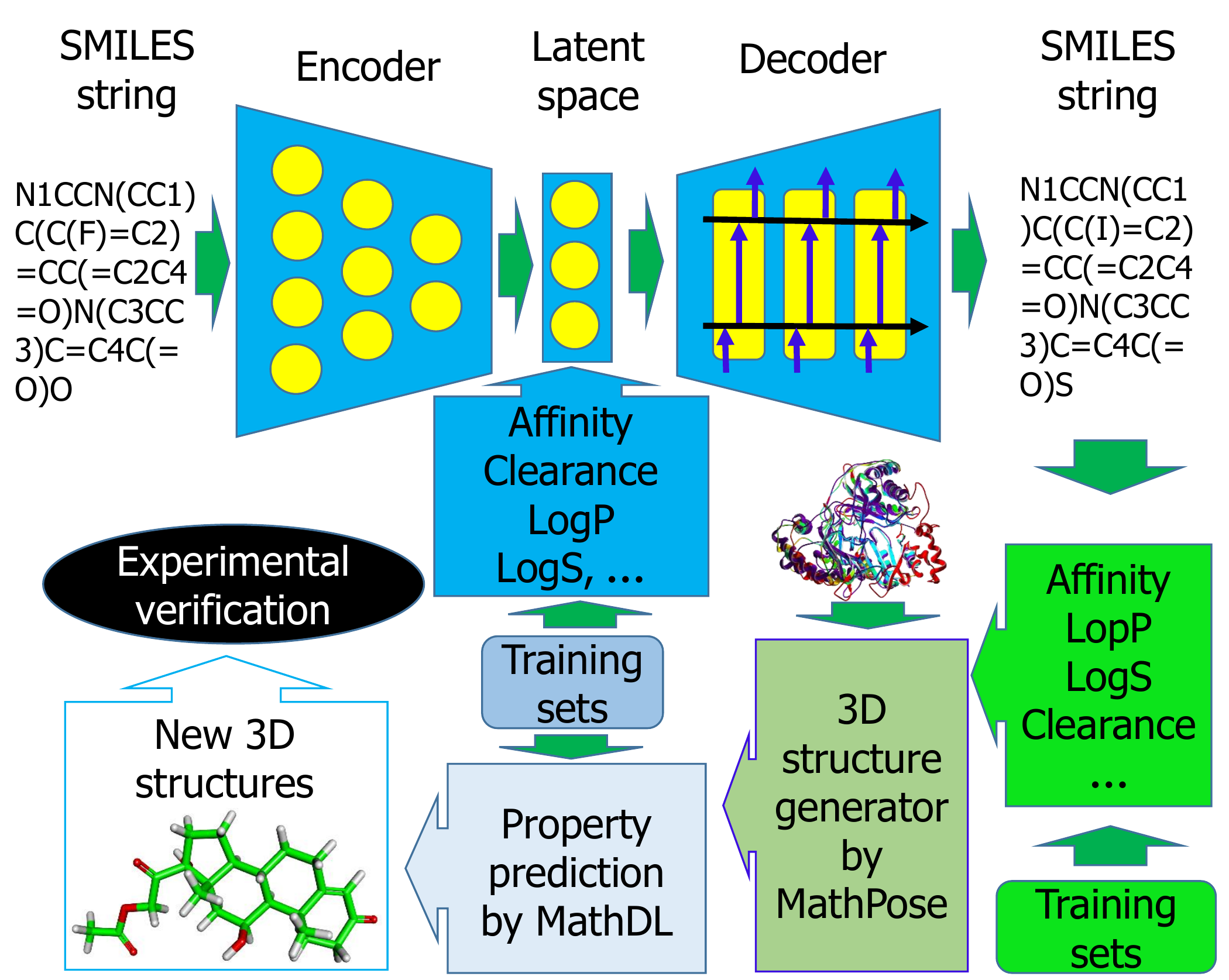}
	\caption{A schematic illustration of  a generative network complex. It consists of an autoencoder that takes SMILES strings (SS) into a drug-property regulated  latent space, a regulated latent space, a LSTM-based autodecoder, a multitask network for the evaluation of binding affinity, partition coefficient (LogP), solubility (LogS), clearance, etc., a 3D structure generator named MathPose, and MathDL, a refined 3D multitask druggable property predictor based on algebraic topology, differential geometry, and graph theory, to select new  drug candidate structures. }
	\label{fig:gan-network}
\end{figure}
In the proposed GNC, the first component is a generative network including encoder, drug-property regulated  latent space, and decoder models. The generative network will take a given SMILES string as input to generate a novel one. The newly generated SMILES strings will be fed into the second component of our GNC, a 2D fingerprint-based deep neural network (2DFP-DNN), so that only ones with desired druggable properties are kept. The next component is the MathPose model which is used to predict the 3D structure information of the compounds selected by 2DFP-DNN. The bioactivities of those compounds are again estimated by the structure-based deep learning model named MathDL. The druggable properties predicted by this last component of our GNC are used as an indicator to select the promising drug candidates. The outline of the GNC is illustrated in Figure  \ref{fig:gan-network}.

\subsubsection{Autoencoder}

An  autoencoder is a type of artificial neural network used to encode a set of data into vectors in the latent space. An autoencoder is typically combined with a decoder to transform the encoded vectors back into SMILES strings.
In the present work, we propose a  latent space technique which controls or regulates various drug properties, such as binding affinity, solubility (LogS), partition coefficient (LogP), clearance, etc.

\begin{figure}[h]
    \centering
    \includegraphics[width=0.45\textwidth]{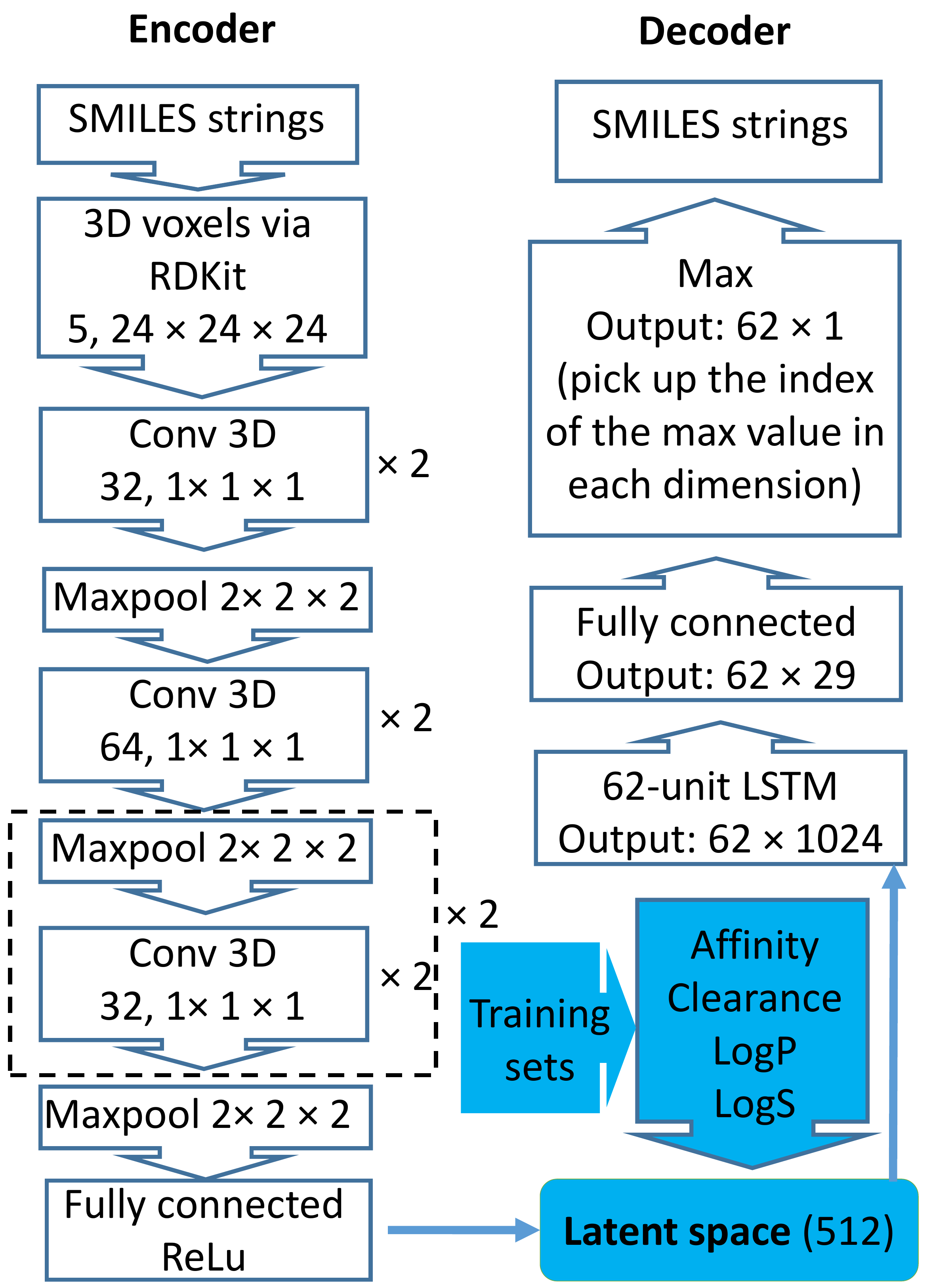}
    \caption{  Illustration of an autoencoder, which consists of a CNN-based encoder, a regulated latent space, and a LSTM-based decoder.}
    \label{fig:enocder-decoder}
\end{figure}

\paragraph{Encoder}

The encoder network in the present work is a convolutional neural network (CNN) which takes
converts SMILES  strings into 3D molecular images before encoding their into the latent space. It consists of  five 3D-convolutional layers. {The number of output channels for each layer are 32, 32, 64, 64, 32, 32, 32, and 32 with kernel sizes all (1, 1, 1), respectively.} For the sake of visualization, the  encoder's architecture is outlined in Figure \ref{fig:enocder-decoder}

In the present  encoder, for each SMILES string, 3D conformers were generated via RDKit \cite{landrum2006rdkit} and optimized using the MMFF94 force field \cite{halgren1996merck} with default settings. {Molecule atoms were then voxelized into a discretized 1 \r{A} cubic grid with sides of length 24 \r{A} prior to a random rotation and 2 \r{A} translation of the molecule. The voxelized value is determined by its atom type and the distance $r$ between neighboring atoms and its center:}
\begin{equation}
	n(r) = 1-\exp[-(r_{\rm vdW}/r)^{12}],
\end{equation}
where $r_{\rm vdW}$ is the van der Waals radius.

{The voxelized values of five types of properties are calculated: hydrophobic, aromatic, H-bond donors, H-bond acceptors, and heavy atoms, leading to five different channels \cite{skalic2019shape}.}

\paragraph{Latent space}

We propose three  latent space regulation  schemes, i.e.,    randomized  output,  controlled output, and optimized output,    to construct new compounds. First,  to generate new compounds from seeds, random noise can be added to the latent space. In other words, the encoded latent vectors can be perturbed by standard Gaussian noise, rendering a possible new latent representation.  The resulting latent vector will be fed into the decoder network.

Additionally, a more interesting control procedure is to select the latent space output through a druggable property assessment. As shown in Figures \ref{fig:gan-network} and \ref{fig:enocder-decoder}, we use the trained encoder to generate latent-space representations of a dataset of interest, such as the BACE dataset. Based on these representations of the dataset and its labels, we train deep learning network models to evaluate and predict various druggable properties, including binding affinities, solubility, partition coefficient, clearance, toxicity, etc. In certain situations, we also build multitask deep learning models to enhance latent-space evaluations. In this approach, each new compound in its latent space representation is evaluated for its druggable properties to determine whether it is to be fed into the decoder.

Finally, a more effective optimization scheme is to actively build new drug candidates in the latent space representation with desirable properties  as shown in  Figures \ref{fig:gan-network} and \ref{fig:enocder-decoder}. With appropriate training datasets, we first construct $m$ latent-space predictive machine learning models as described above for $m$ different properties, such as binding affinities, solubility, partition coefficient, clearance, etc.  For each property, we set up a target value, $y_{j0}$. We then build  an $L_2$ loss function to optimize a given $n$-component latent-space vector $X\in {\mathbb R}^n$:
 \begin{equation}\label{latent}
   \min \sum_{j=1}^m k_j(\hat{y}_j(X)-y_{j0})^2,
\end{equation}
where $k_j$ is a preselected weight coefficient for the $j$th property and $\hat{y}_{j}(X): {\mathbb R}^n\rightarrow {\mathbb R}$ is the predicted $j$th  property value of the latent-space vector $X$  from latent-space machine learning models.  Alternatively, we also use other metrics, such as $L_1$ or mixed metrics for constructing the loss function. The optimization with a gradient decent algorithm leads to an iterative scheme for regularizing the latent-space vector $X$.  Alternatively, a Monte Carlo procedure can be implemented.

Target values $y_{j0}$ can be chosen to optimize potential drugs. In case of binding affinity (BA), we use a targeted value of $y_{\rm BA}\le -9.6 $kcal/mol. For  LogP, we set $y_{\rm LogP}\le 5$. Note that additionally constraints, such as, similarity,  Lipinski's  rule of five  \cite{lipinski1997experimental} or their variants for   druglikeness can be easily implemented with Eq. (\ref{latent}).

\paragraph{Decoder}
The decoder network here consists of several LSTMs.  LSTMs are variants of RNNs that were proposed to handle language processing problems, which require the network to take into account the relationships between words rather than simply interpreting each word independently. RNNs are designed to pass fixed-size pieces of information from one neuron to others in the network. However, RNNs are not very effective at processing information with long-term dependencies, as the persistence of information within the network is somewhat short-lived. As a result, LSTMs were designed to overcome this problem \cite{hochreiter1997long, jozefowicz2016exploring, johnson2017google}. The  encoder network was trained in the shape  encoder framework.

In each LSTM unit, there is a cell consisting of an input gate, an output gate, and a  forget gate which are described in the following equations
\begin{equation}
    H_t=o_t*\operatorname{tanh}(C_t),
\end{equation}
where $o_t$ depends on its input $X_t$ and the output of the last layer $H_{t-1}$:
\begin{equation}
    o_t=\sigma(W_o[H_{t-1},X_t]+b_{o}).
\end{equation}
Here, $\sigma$ is the activation function.  Now, the cell state at the $i$th layer is given by:
\begin{equation}
    C_t=f_t*C_{t-1}+i_t*\hat{C}_t,
\end{equation}
where $\hat{C}_t$ is the change of the cell state at the $i$th layer
\begin{equation}
    \hat{C}_t=\operatorname{tanh}(W_C[h_{t-1},x_t]+b_C),
\end{equation}
and $f_t$ and $i_t$ are given by the following,
\begin{align}
    f_t &= \sigma(W_f[H_{t-1},X_t]+b_f) \\
    i_t &= \sigma(W_i[H_{t-1},X_t]+b_i).
\end{align}
The updated cell state, $C_t$, is then passed to the next layer along with the output of the current layer and includes accumulated information from all previous layers so that the network can handle long-term dependencies between inputs.

The purpose of LSTMs in our case is to decode molecules from the encoded vectors in the latent space. There is some variation in the decoding process via the use of probabilistic sampling. Due to the LSTM's ability to handle long-term dependencies, it can learn SMILES grammar, and build SMILES strings by selecting the next token proportionally to its predicted probability \cite{skalic2019shape}. This means that some variation from the seed SMILES string will occur. As a result, even when the input is the same, the output will not always be  the same. This causes the generated SMILES strings to be different from their seeds. Our LSTM  decoder has 5 layers as the same as the number of layers of the aforementioned CNN  encoder. The architecture of the  decoder is depicted in Figure \ref{fig:enocder-decoder}.  The Adam optimizer was applied with the learning rate 0.001 and a batch size of 128 to minimize the following loss
\begin{equation}
    L = -\frac{1}{N}\sum_{i=1}^N\sum_{j=1}^M y_{j}^{(i)}\log{p_{j}^{(i)}},
\end{equation}
where $y_{j}^{(i)}$ and $p_{j}^{(i)}$, respectively, represent the  the ground-truth and the predicted probability for component $j$th in the $i$th SMILES string. Also,  $N$ is the number of samples in each batch and $M$ is the length of the SMILES string.

\subsubsection{2D fingerprint-based binding affinity predictors (2DFP-DNN)}\label{sec:DNN}

The predictors are deep neural networks (DNN) pre-trained on our own training sets. A DNN mimics the learning process of a biological brain by constructing a wide and deep architecture of numerous connected neuron units. A typical DNN often includes multiple hidden layers. In each layer, there are hundreds or even thousands of neurons. During the learning stage, weights on each layer are updated by backpropagation. A   deep neural network is able to construct hierarchical features and model complex nonlinear relationships.

The purpose of our DNN predictors is to predict the binding affinities and other properties of the generated compounds and, based on that, screen ideal drug candidates meeting our criteria. Binding affinity assesses a drug's binding strength to its target, which is one of the most important drug properties \cite{parenti2012advances, gao2015binding}. The input of predictor networks is 2D molecular fingerprints. In our case, a combination of ECFP \cite{rogers2010extended} and MACCS \cite{durant2002reoptimization} fingerprints were used, yielding 2214 bits of features (2048 bits from ECFP and 166 bits from MACCS) in total. The output of the network is the drug properties, such as binding affinity, log P, and log S. During the training and prediction processes, the SMILES strings of compounds were first transformed to their 2D fingerprints and then fed into the network. The fingerprint transformation from SMILES strings was conducted by RDKit \cite{landrum2006rdkit}.

With appropriate training data, we can construct multitask DNNs for simultaneous predictions of binding affinity, log P, log S, and toxicity \cite{wu2018topp,wu2018quantitative}. Our DNN predictor networks have 4 layers with 3000, 2000, 1000, and 500 neurons in each hidden layer, respectively. For training, we used stochastic gradient descent with a momentum of 0.5. We trained each network for 2000 epochs with a mini-batch size of 4. We used a learning rate of 0.01 for the first 1000 epochs and reduced it to 0.001 for the last 1000 epochs. Our tests indicate that adding dropout or $L_2$ decay does not necessarily increase the accuracy of the networks, and as a consequence, we omitted these two techniques. The DNN training and prediction are performed by Pytorch \cite{paszke2017pytorch}.

\subsubsection{MathDL for energy prediction}\label{sec:MathDL}
\begin{figure}[h]
    \centering
    \includegraphics[width=0.8\textwidth]{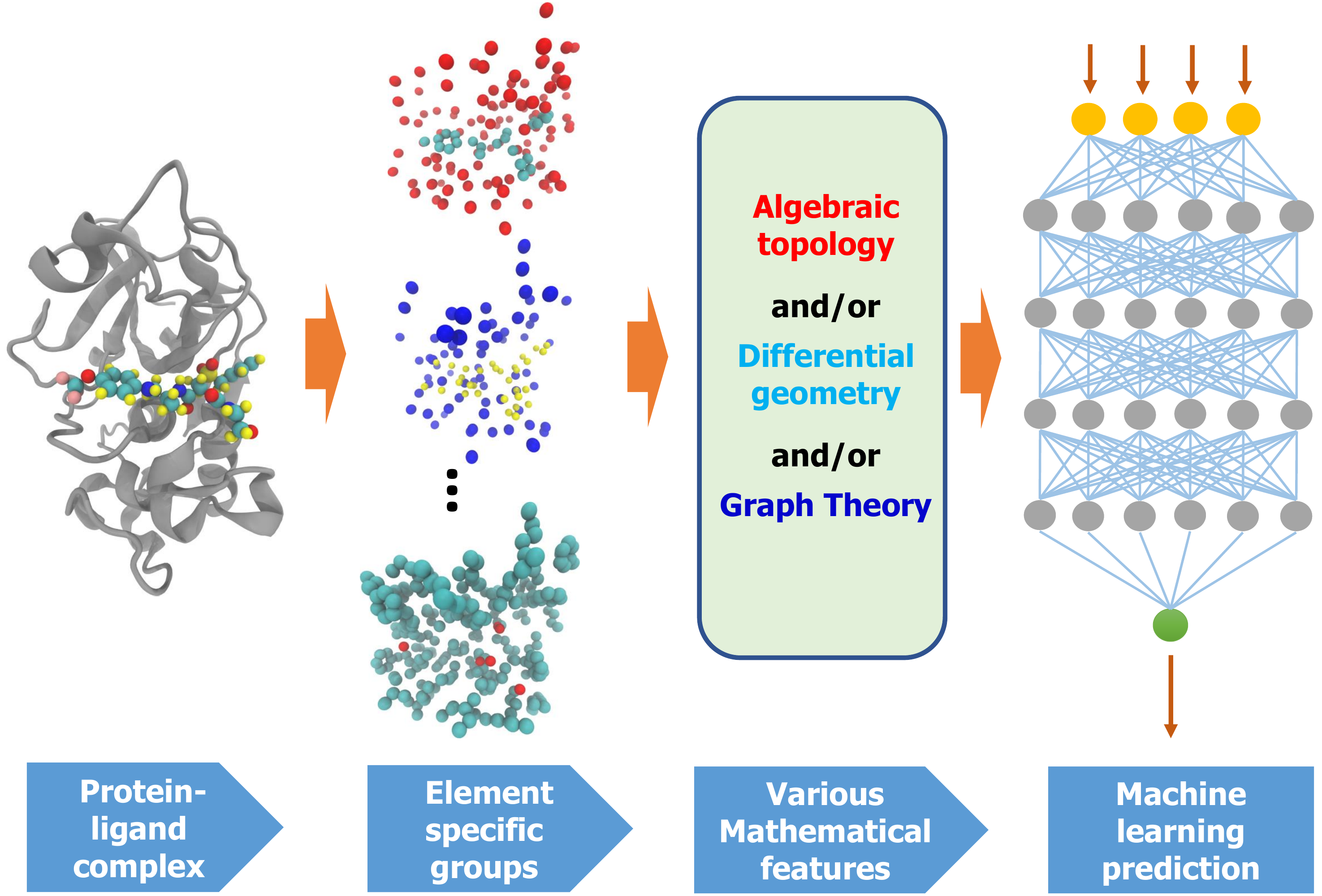}
    \caption{A schematic illustration of the MathDL for binding affinity prediction in which the combination of several advanced mathematical representations is integrated with sophisticated CNN models.}
    \label{fig:MathDL}
\end{figure}
Our MathDL is constructed by  the integration of mathematical representation features and deep learning networks to generate a powerful binding affinity predictor \cite{nguyen2019mathematical,nguyen2019D3R}. Specifically, the MathDL is the blend of intensively validated models based on algebraic topology \cite{cang2015topological,cang2018representability,cang2018integration}, differential geometry \cite{nguyen2019dg}, and graph theory \cite{nguyen2017rigidity,nguyen2019agl}. In these methods, algebraic topology model makes use of persistent homology in multi-component and multi-level manners to characterize protein-ligand complexes by topological invariants, i.e., Betti numbers counting various dimensional holes. In the 3D space, we have Betti-0, Betti-1, and Bett-2 which receptively counts the numbers of independent components, recognizes numbers of rings, and  accounts for the cavity information \cite{zomorodian2005computing,xia2014persistent,xia2015persistent}. Our previous work, we have shown that algebraic topology network has outperformed other state-of-the-art methods in  the classifying proteins \cite{cang2015topological} and active/inactive compounds \cite{cang2018representability}, and the predictions of protein-ligand binding affinity  \cite{cang2017topologynet, cang2018representability},  toxicity \cite{wu2018quantitative}, lop P, and log S \cite{wu2018topp}.

Differential geometry describes how molecules assume complex structures, intricate shapes and convoluted interfaces between different parts.  \cite{wei2010differential}
In our differential geometry-based model,  essential chemistry, physical, and biological information are encoded into the low-dimensional interactive manifolds which are extracted from high-dimensional data space via a multiscale discrete-to-continuum mapping \cite{xia2015multidimensional,nguyen2019dg}. Thereby, the molecular structures and atomic interactions can be conveniently represented via interactive curvatures, interactive areas, etc. Numerous numerical validations have shown that the differential geometry model has achieved the state-of-the-art performances on various biological prediction tasks, namely drug toxicity, molecular solvation, and protein-ligand binding affinity \cite{nguyen2019dg}.

Recently, we have developed a powerful algebraic graph-based scoring function which encodes the important physical and biological properties such as hydrogen bonds, hydrophilicity, hydrophobicity, van der Waals interactions, and electrostatics from the high-dimension space into the low-dimension description via the invariants extracted from Laplacian, its pseudo-inverse, and adjacency matrices \cite{nguyen2019agl}. Algebraic graph theory-based models have been widely utilized in the study of physical modeling and molecular analysis such as  chemical analysis \cite{plavvsic1993relation, janezic2015graph}, protein flexibility analysis \cite{go1983dynamics, atilgan2001anisotropy, bahar2010global}. Despite  its popularity, the graph-based quantitative models typically are not as competitive as other quantitative approaches due to no categorization on element types and the missing crucial non-covalent interactions. This missing information has been encoded in multiscale weighted colored subgraphs in our newly designed algebraic graph-based model, named AGL-Score. Extensive numerical validation on PDBbind benchmarks with various evaluation metrics, namely scoring power, ranking power, docking power, and  screening power has shown that our AGL-Score has outperformed other state-of-the-art methods on these evaluations which are the standard criteria for virtual screening in drug discovery \cite{nguyen2019agl}.

The combination of these three powerful  models gives rise to the MathDL model which is expected to be one of the most accurate binding affinity predictors available in the literature. Indeed, the MathDL model achieved the top performances on the affinity ranking and free energy prediction for Cathepsin S (CatS) inhibitors in the Drug Design Data Resource (D3R) Grand Challenge 4 (GC4), a worldwide competition series in the computer-aided drug design\cite{nguyen2019D3R}. Also, MathDL model was the competitive scoring functions on the binding energy predictions for beta-secretase  1 (BACE) compounds in GC4 \cite{nguyen2019D3R}. The outline of the MathDL model is depicted in Figure \ref{fig:MathDL}.

\subsubsection{MathPose for 3D structure prediction}\label{sec:MathPose}

\begin{figure}[h]
    \centering
    \includegraphics[width=0.7\textwidth]{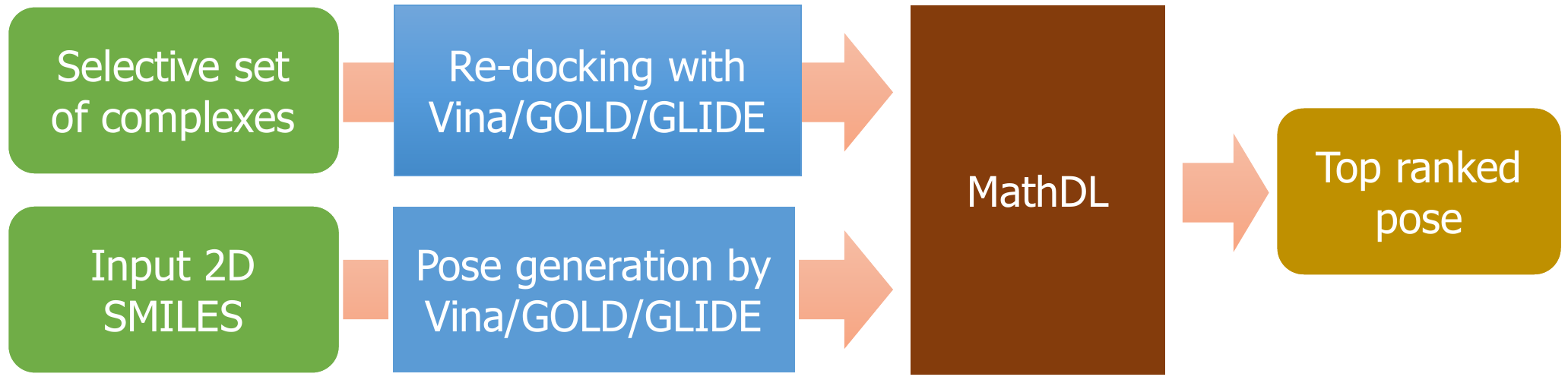}
    \caption{A schematic illustration of the MathPose approach for 3D structure generation from a given input 2D SMILES string.}
    \label{fig:MathPose}
\end{figure}
In our recent work, we have successfully designed an AGL-Score model to achieve the best performances in docking power metrics which validate the scoring function's ability to identify the ``native pose'' from the computer-generated poses \cite{nguyen2019agl}. Specifically, on the CASF-2007 benchmark, \cite{cheng2009comparative} our AGL-Score achieves 84\% accuracy on the docking power assessment \cite{nguyen2019agl}. The second best scoring function on this benchmark is from GOLD software with ASP fitness score (82\%)  \cite{cheng2009comparative}. Our scoring function is still the top performer on docking power test of CASF-2013 benchmark \cite{li2014comparative} with the accuracy as high as 90\%\cite{nguyen2019agl}, followed by the machine learning based-scoring function $\Delta_{\rm vina}{\rm RF}_{20}$ (87\%)\cite{wang2017improving}. With such promising results, it is expected that the replacement of the single AGL-Score model by intricate MathDL scoring function will certainly improve the quality of pose ranking. This results in the MathPose model whose framework is outlined in Figure \ref{fig:MathPose}. In our MathPose, besides the SMILES string of the interested ligand $L$, we select a set of complexes having similar binding sites to the one the ligand $L$ can bind to. A pool of nearly 1000 poses for the ligand $L$ is generated by several common docking software, namely Autodock Vina \cite{trott2010autodock}, GOLD \cite{jones1997development}, and GLIDE \cite{friesner2004glide}. Additionally, three docking software packages are utilized to re-dock the complexes in the selective data set to form at least 100 decoy complexes per input. Then, our MathDL will be trained on these decoy sets to learn the calculated root mean squared deviation (RMSD) between the decoy and native structures. The trained MathDL will be applied to pick up the top-ranked pose for the given ligand $L$.

\subsection{The analysis of generated compounds}

\paragraph{The 2D similarity analysis between generated compounds and their seeds.}
 To investigate how ``novel'' our generated compounds are from their seeds, a similarity analysis was performed on them. The 2D molecular SMILES strings of the generated molecules were also transformed into 2D molecular fingerprints and then the similarity scores between the fingerprints of the generated molecules and their seeds were calculated. The fingerprints were the same ones used in the DNN predictors, a combination of ECFP and MACCS molecular fingerprints. The criteria used for the similarity scores was the Tanimoto coefficient \cite{bajusz2015tanimoto}. The fingerprint transformation was also conducted by RDKit \cite{landrum2006rdkit}.

\paragraph{The k-means clustering analysis of generated compounds.}
Cluster analysis or clustering is the task of grouping a set of objects in such a way that objects in the same group are more similar to each other than to those in other groups. It has already been widely applied to protein conformation analysis. \cite{gao2015molecular, gao2016network, gao2017network}
To present the diversity of our generated active compounds, k-means clustering analysis was performed. The input features were the same molecular fingerprints discussed above, and the k-means clustering was conducted by scikit-learn \cite{pedregosa2011scikit}. For each cluster, the center was extracted to represent the cluster.


\section{Results}

To examine and validate the performance of our proposed GNC  for generating new compounds for drug targets, we consider two specific targets, namely Cathepsin S (CatS) set and Beta-Secretase 1 (BACE). These two targets appeared in the D3R Grand Challenges, worldwide competition series in computer-aided drug design \cite{gaieb2019d3r,nguyen2019D3R}, with components addressing pose-prediction, affinity ranking, and free energy calculations.

Both   CatS and BACE are potential targets for significant human diseases. CatS constitutes an 11-member family of proteases involved in protein degradation. It is highly expressed in antigen-presenting cells, where it degrades major histocompatibility complex class II (MHC II)-associated invariant chain. CatS is a candidate target for regulating immune hyper-responsiveness, as the inhibition of CatS may limit antigen presentation \cite{ameriks2010diazinones,wiener2010thioether}. BACE is a transmembrane aspartic-acid protease human protein encoded by the BACE1 gene. It is essential for the generation of beta-amyloid peptide in neural tissue \cite{vassar2009beta}, a component of amyloid plaques widely believed to be critical in the development of Alzheimer's, rendering BACE an attractive therapeutic target for this devastating disease \cite{prati2017bace}. The rest of this section is devoted to the utilization of the proposed GNC on the exploration of new potential drugs for CatS and BACE targets.

\subsubsection{Faithful validation of generative network on CatS and ZINC data sets}

To assess the performance of the autoencoder on the CatS data set, we converted all SMILES strings in the data set into the canonical form using RDKit \cite{landrum2006rdkit}, and kept only the strings with length no more than 60. This was done because the decoder network was designed to produce only SMILES strings of length at most 60. This left us with 1858 of the 2847 molecules in the CatS training set. After feeding these molecules through the network, 1427 (76.8\%) yielded valid SMILES strings, with none being identical to the original.

We also tested the performance of the autoencoder on a larger data set of 1 million molecules, randomly chosen from the same subset of the ZINC 15 \cite{sterling2015zinc} data set from which the training samples were drawn. The training set produced from the ZINC 15 data set contains 192,813,983 molecules, 26,880,000 of which were previously seen by the autoencoder during training. From these 1 million molecules, 994,219 (99.4\%) yielded valid SMILES strings, and 2,724 (0.27\%) SMILES strings were reproduced exactly. A high valid molecule generation rate and a low reconstruction rate enable us to generate meaningful compounds with highly diverse chemical properties.

\subsection{BACE}

\subsubsection{Data preparation}
\begin{figure}[h]
    \centering
    \includegraphics[width=1.0\textwidth]{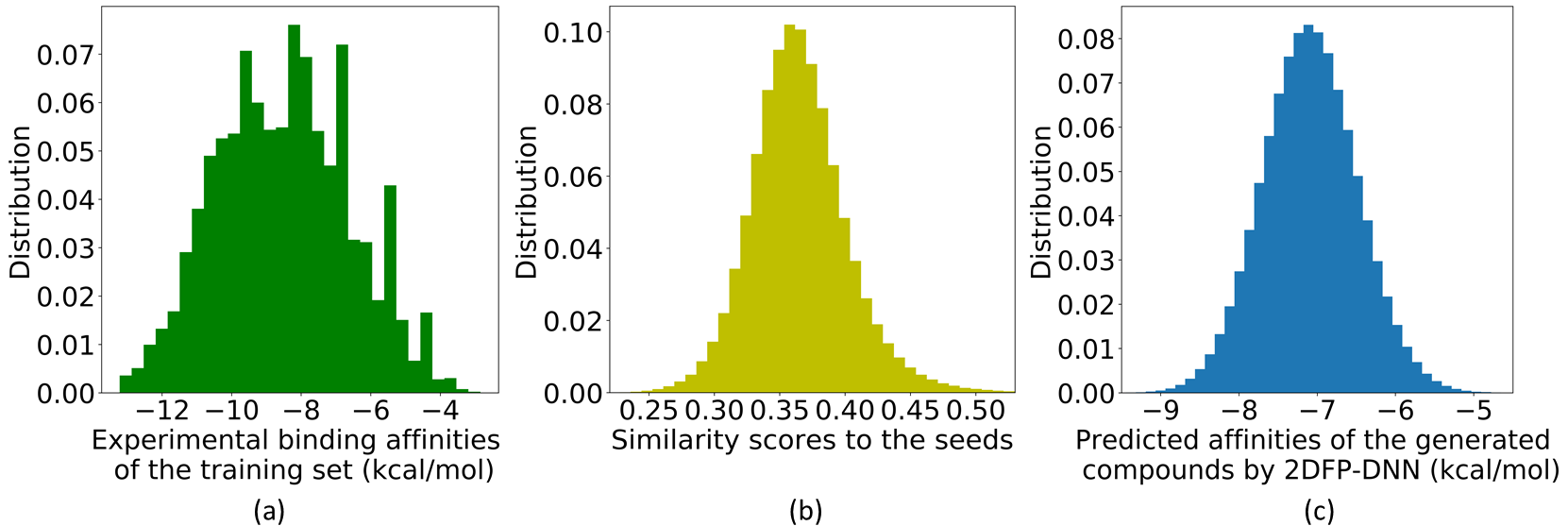}
    \caption{Distributions on the BACE set. (a) The distribution of the experimental binding affinities in the BACE training set; (b) The similarity distribution of new molecules compared with their seeds in the BACE set. (c) The distribution of new BACE molecules'  binding affinities predicted by 2D fingerprint network model 2DFP-DNN.}
    \label{fig:dist-bace}
\end{figure}
To enable the proposed GNC to generate meaningful BACE inhibitors, one needs to supply it with seed molecules closely related to the BACE target. To this end, we combine all BACE inhibitors provided in the D3R Grand Challenge 4 (\url{https://drugdesigndata.org/about/grand-challenge-4}) with the BACE ligands having the reported binding affinity on the ChemBL database (\url{https://www.ebi.ac.uk/chembl/}). That results in a BACE data set of 3916 compounds with binding affinities ranging from -2.84 kcal/mol to -13.22 kcal/mol. If one sets -9.56 kcal/mol as a threshold to label a compound as active, that BACE data set has 1231 active ligands. The distribution of binding affinity in the BACE data set is shown in Figure \ref{fig:dist-bace}a. That figure reveals that most of the molecules in our collected data set having affinities between -10 kcal/mol and -7 kcal/mol. Also, there are more BACE inhibitors with binding strength less than -10 kcal/mol than ones having binding affinity higher than -6 kcal/mol.

\subsubsection{Structure generation}

\begin{figure}[!htb]
    \centering
    \includegraphics[width=.7\textwidth]{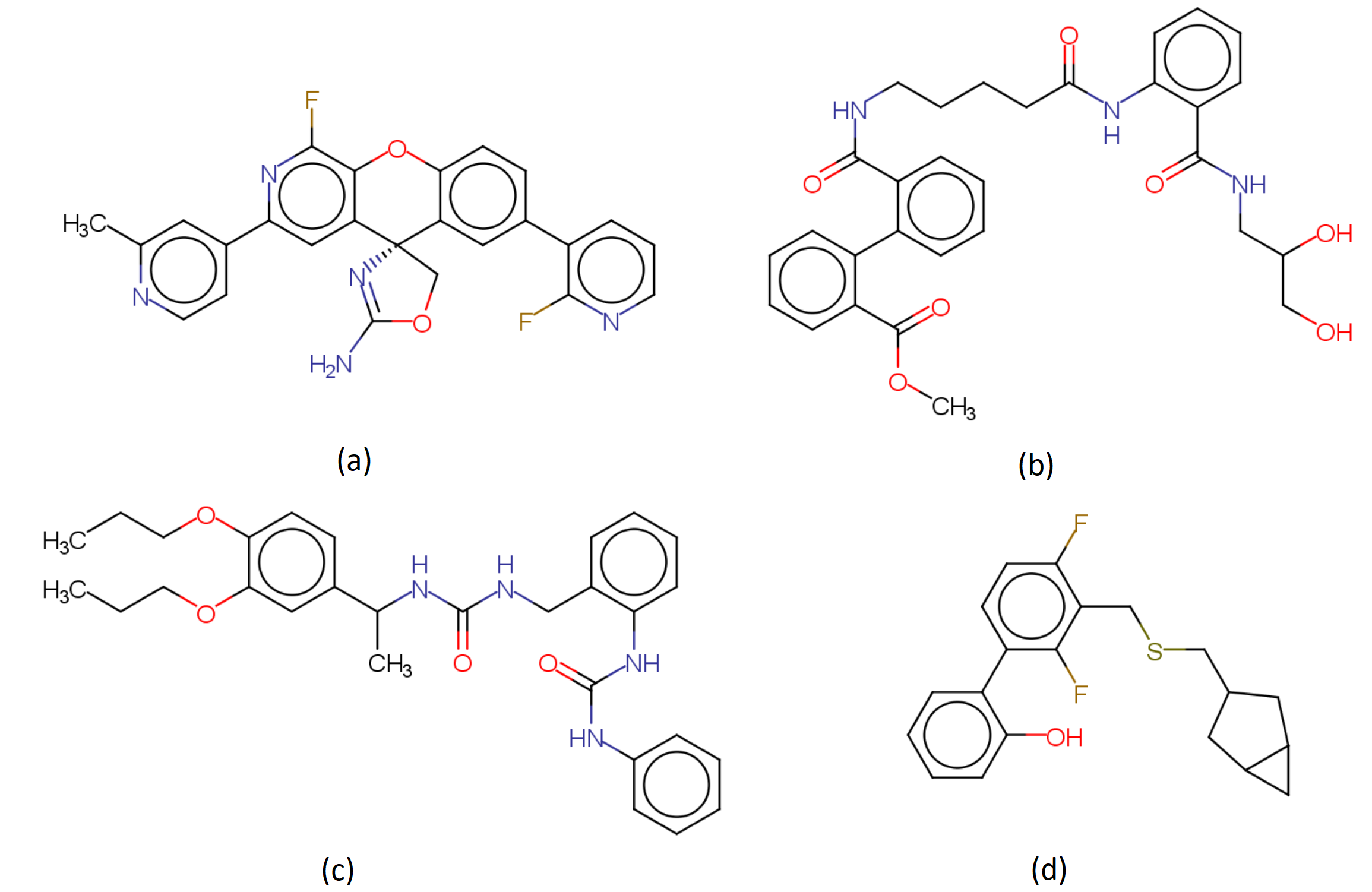}
    \caption{The illustration of similarity between a seed molecule in the BACE set and some generated compounds: (a) The seed; (b) The most similar compound generated from the seed (similarity score=0.50); (c) A compound with a medium similarity score of 0.34; (d) The most different one from the seed (similarity score=0.23). }
    \label{fig:mol-sim}
\end{figure}
By feeding the BACE data set of 3916 compounds to the generative network, as many as 2.8 million valid compounds were generated by supercomputers in less than one week.
To indicate how ``novel'' our generated compounds are from their seeds, the similarity score between each generated compound and its seed is calculated. The similarity score distribution is illustrated in Figure \ref{fig:dist-bace}b.  It is  revealed from the figure that the similarity scores of the generated compounds have a broad range varying from 0.15 to 1.00. This means that our generated compounds cover a very large chemical space. A similarity score being 1 indicates the generated compound is exactly the same as the seed. Fortunately, this is very rare, happening only 9 times in all 2,727,379 generated compounds. In most cases, the similarity scores are very low with an average value of 0.34, implying the wide range of diversity among the generated samples.

To further verify that the generated compounds are really   unique from the seeds, a seed molecule and several generated compounds are shown in Figure \ref{fig:mol-sim}. In which, Figure \ref{fig:mol-sim}a depicts a seed, Figure \ref{fig:mol-sim}b illustrates the most similar compound generated from the seed, Figure \ref{fig:mol-sim}c plots a compound with a medium similarity score of 0.34, and Figure \ref{fig:mol-sim}d presents the most different one.
One can realize that even the most similar one with a similarity score as high as 0.50, chemical structures are still quite different due to the replacement of the fused ring by a carbon chain (see Figures \ref{fig:mol-sim}a and \ref{fig:mol-sim}b).

\subsubsection{Binding affinity screening by 2DFP-DNN}\label{sec:BACE_dist_affinity}

To efficiently select the potential drug candidates, we carry out the 2D fingerprint DNN model discussed in Section \ref{sec:DNN} to predict the binding affinities of more than 2.7 millions compounds. Figure \ref{fig:dist-bace} illustrates those predicted energies. From Figure \ref{fig:dist-bace}, one can notice that the predicted affinities of the generated BACE compounds are distributed in a Gaussian manner. This result is probably due to the Gaussian-like distribution of similarity scores between generated ones and their corresponding seeds depicted in Figure \ref{fig:dist-bace}b.

The range of their binding affinities of predicted molecules is widely spread from -3.89 kcal/mol to -10.20  kcal/mol, confirming that large chemical space is covered.
The peak is at -7.1 kcal/mol, which means about half of the generated compounds have binding affinity smaller than -7.1 kcal/mol. Among this first half with the binding affinity smaller than -7.1 kcal/mol, 5 compounds have predicted binding affinity smaller than -10 kcal/mol which indicates they are promising drug candidates. Moreover, there are 2130 compounds with binding affinity smaller than -9 kcal/mol, and 178250 compounds with the binding affinities smaller than -8 kcal/mol. In this work, we use a common binding affinity threshold, i.e. -9.56 kcal/mol, to screen out high-likely less active compounds. As a result, we are left with 99 generated inhibitors having the lowest binding energy in term of kcal/mol.

It is noticed that the 2D fingerprint DNN model for binding affinity prediction only relies on the ligand information without the involvement of target proteins. Therefore, its accuracy is not as high as its 3D counterparts (e.g. MathDL) \cite{cang2018representability,nguyen2019D3R} in which the interactions between the target binding site and the interesting compounds are fully incorporated. However, it is projected to be time consuming when carrying out those 3D-based binding affinity predictor models on a large pool of molecules. Thus, in this work, we make use of the advantage of simple calculations in the 2D-based models to filter out a large number of compounds with highly predicted affinities.

\subsubsection{Clustering analysis of selected compounds}
 \begin{figure}[!htb]
    \centering
    \includegraphics[width=.8\textwidth]{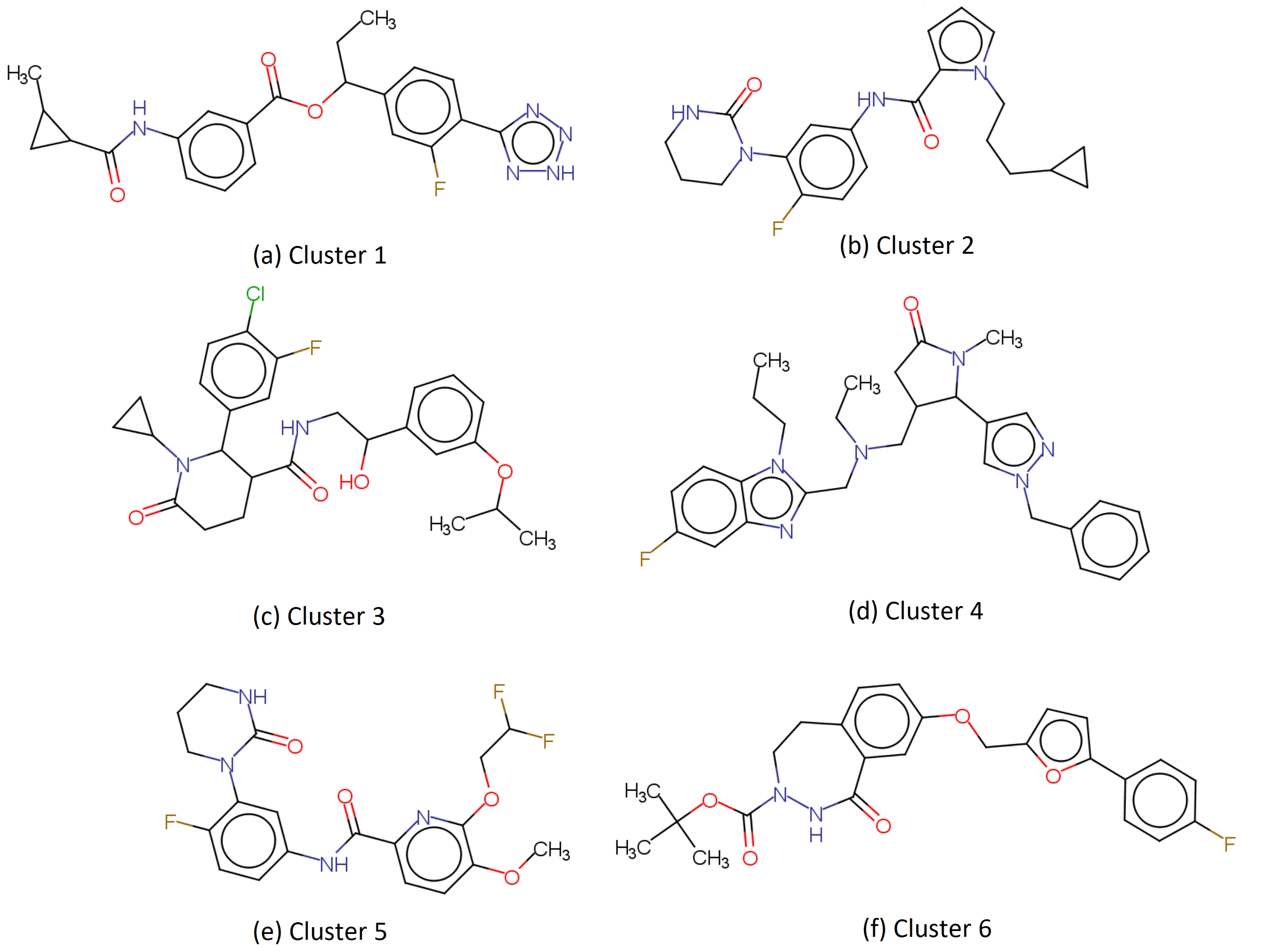}
    \caption{The center of the 6 clusters found in our BACE generated set.}
    \label{fig:clusters}
 \end{figure}
To illustrate how diverse our generated active compounds are, clustering analysis was performed on the 99 generated compounds with the most highly predicted binding affinities discussed in Section \ref{sec:BACE_dist_affinity}. By carrying out k-means clustering method, one can find 6 clusters in our generated set, and the center of each cluster is shown in Figure \ref{fig:clusters}.

Statistically, the sizes of these 6 clusters are 7, 38, 10, 7, 12, and 25, respectively. Inside these 6 clusters, the average similarity scores to the centers are 0.69, 0.58, 0.62, 0.66, 0.63, and 0.67, respectively, which indicates the compounds in the same cluster are relatively similar. By contrast, the similarity scores between different clusters are much lower. Specifically, the similarity score between these 6 cluster centers are only around 0.40; thereby, implying the high diversity in our generated compounds.

Among these 6 clusters, cluster 2 is the biggest one with 38 compounds. Moreover, it contains the largest numbers of the highly predicted binding affinities. Particularly, cluster 2 has 5 compounds with predicted binding affinities smaller than -10 kcal/mol. Since the compounds in the same cluster are similar, it suggests that other compounds in cluster 2 may also have a high potential to become drugs. SMILES strings of all 99 compounds in 6 clusters are included in Table S1 in Supporting Information.

\subsubsection{Binding affinity screening by MathDL}
 \begin{figure}[!htb]
    \centering
    \includegraphics[width=0.7\textwidth]{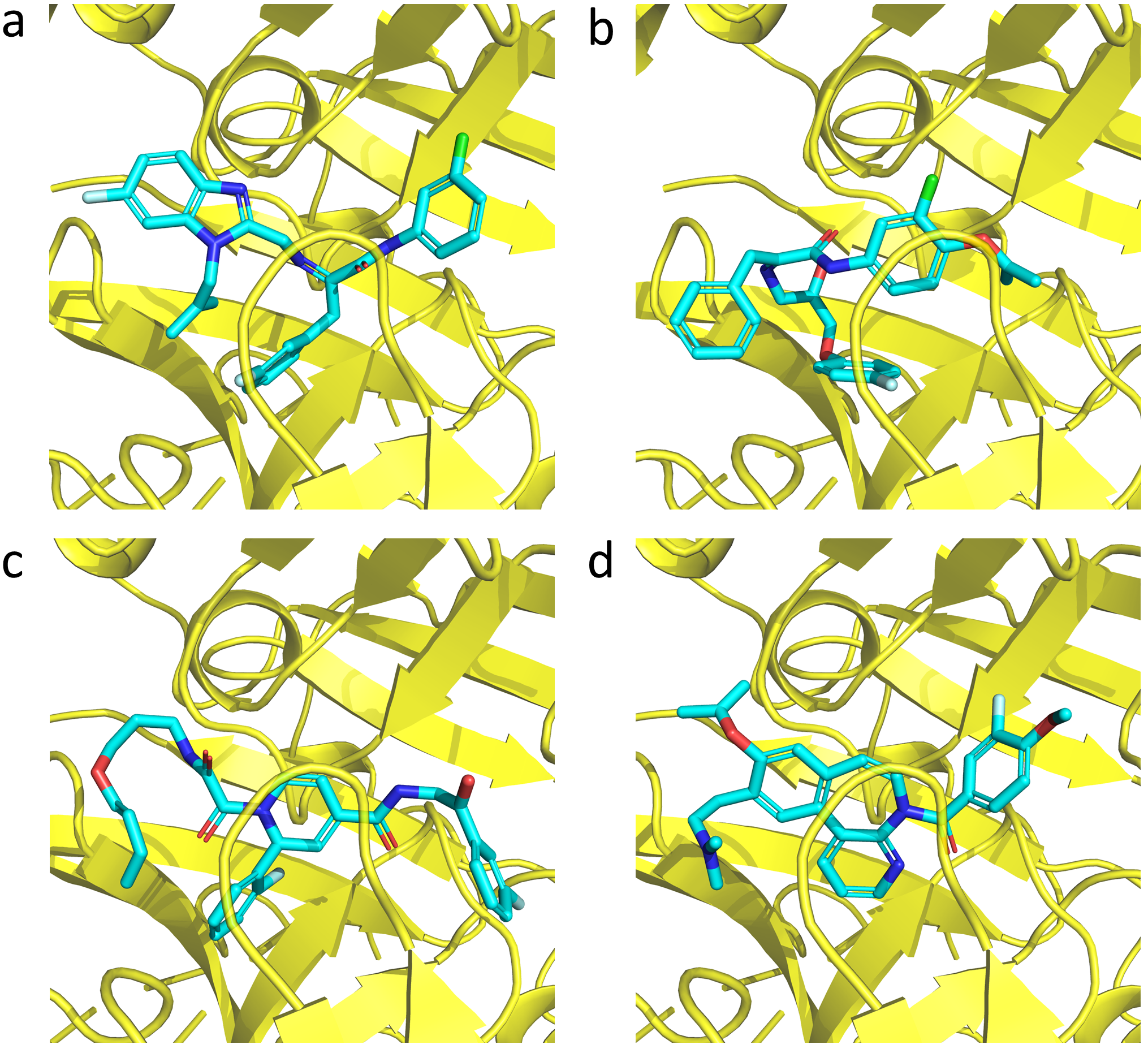}
    \caption{Top 4 generated BACE compounds having the lowest binding affinities predicted by MathDL model. Their 3D structures were constructed by MathPose. Their IDs under our naming system and the predicted energies are, respectively (a) BACE\_gen\_35 (-8.263 kcal/mol); (b) BACE\_gen\_66 (-8.258 kcal/mol); (c) BACE\_gen\_29 (-8.202 kcal/mol); and (d) BACE\_gen\_25 (-8.20 kcal/mol). Their SMILES strings are provided in Table S1, and their corresponding 3D structures are included in File S1. All of those molecules were docked to protein extracted from the complex with PDB ID 3dv5.}
    \label{fig:BACE_top4}
 \end{figure}
The DNN model using the 2D fingerprint features, as discussed in Section \ref{sec:BACE_dist_affinity}, only relies on the ligand information and lacks the receptor environment. As a result its reliability is not guaranteed when identifying the most promising drug candidates. It has been shown that structure-based models often outperform the ligand-based models in diverse datasets\cite{cang2018representability, nguyen2019mathematical, nguyen2019D3R}. Therefore, our MathDL model, discussed in Section \ref{sec:MathDL}, is utilized to re-rank the compounds picked out by 2DFP-DNN models. The MathDL model is trained on the BACE data set of 3916 compounds whose 3D structures are generated by MathPose mentioned in Section \ref{sec:MathPose}.

The Kendall's Tau coefficient ($\tau$) and Pearson correlation coefficient ($R_p$)  of the cross-validation on the training data are 0.608 and 0.797, respectively. These accuracy evaluations guaranteed a well-trained MathDL model on that specific training set. A generated compound set of 99 molecules are fed into MathPose to obtain 3D structures provided in File S1 in Supporting Information. All of them were docked to the protein extracted from a complex with PDB ID 3dv5. Their binding affinities are, then, predicted by the aforementioned trained MathDL model. It is noticed that binding affinity values of 99 generated molecules predicted by MathDL are higher than ones estimated by the 2D fingerprint DNN approach in term of kcal/mol. Specifically, based on MathDL predictor, the lowest binding energy is -8.263 kcal/mol, the highest energy is -5.972 kcal/mol, and the averaged energy over 99 compounds is -7.33 kcal/mol. Figure \ref{fig:BACE_top4} illustrates the binding poses of top four ligands, namely BACE\_gen\_35, BACE\_gen\_66, BACE\_gen\_29, and BACE\_gen\_25, in term of affinity. The predicted energies of those top 4 molecules are -8.263 kcal/mol, -8.258 kcal/mol, -8.202 kcal/mol, and -8.20 kcal/mol, respectively.

Despite having nearly the same values of predicted affinities among those top 4 compounds, they are quite different molecules judged by their 2D similarity scores. Specifically, among those 4 compounds, BACE\_gen\_35 and BACE\_gen\_66 are the most similar structures but their similarity score is as low as 0.265. In addition, BACE\_gen\_29 and  BACE\_gen\_25 are the most dissimilar compounds with 2D singularity score being 0.11. Generating very low binding affinity compounds with diverse chemical formulas is an important goal for the pre-clinical stage since that will enhance the chance of selecting promising drug candidates with low risk of having a side effect. Obtaining top and disparate molecules demonstrates the capacity of our proposed GNC in capturing the wide range of chemical space.


\subsection{CatS}

\subsubsection{Data preparation}

\begin{figure}[h]
    \includegraphics[width=1.0\textwidth]{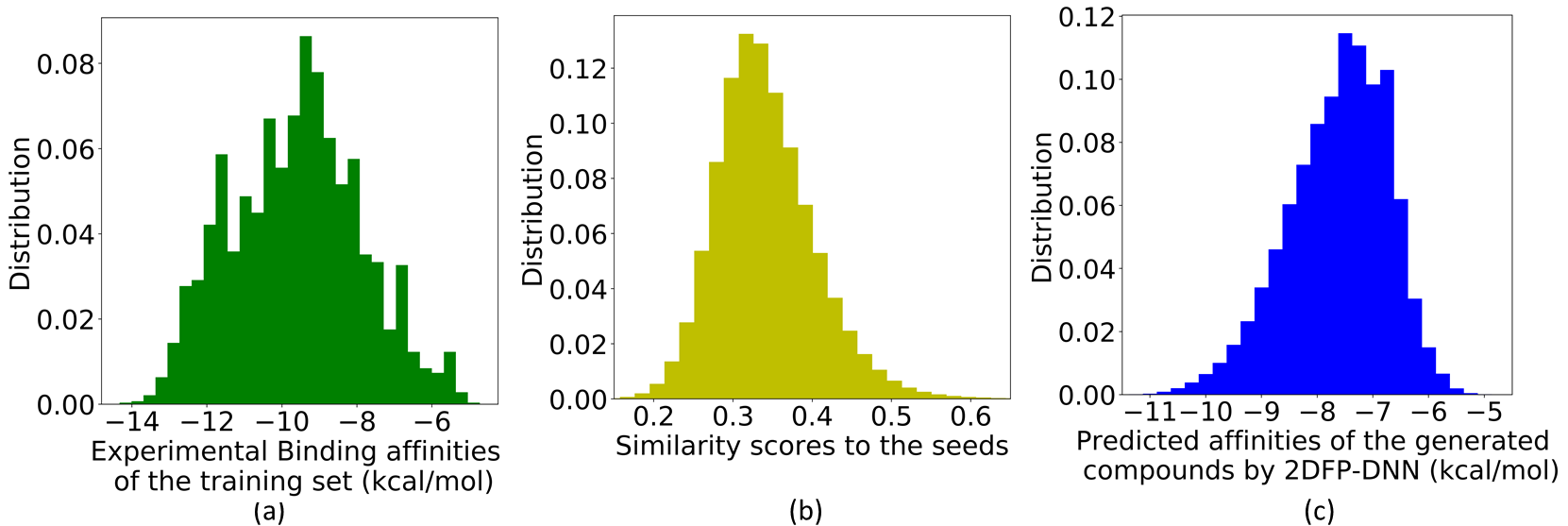}
    \caption{The three distributions about the CatS set. (a) The distribution of experimental binding affinity in the CatS data set; (b) The distribution of similarity scores to their seeds in the CatS generated set. (c) he distribution of the CatS generated set's binding affinities predicted by 2D fingerprint network model 2DFP-DNN.}
    \label{fig:dist-cats}
\end{figure}
Similar to the BACE target, CatS inhibitors were presented in the D3R grand challenges. Thus, these compounds ($n=593$) are used as seeds to produce new CatS molecules.  Other CatS compounds reported in the ChemBL database are also included in our seeds. In total, we collected a data set of 2847 compounds. The binding affinity of these molecules ranges from -4.72 to -14.33 kcal/mol. As with the BACE data set, we chose -9.56 kcal/mol as the threshold for the active compound selection. With this threshold, 1461 of the 2847 compounds in our seeds are active. The distribution of binding affinity in our collected CatS data set is shown in Figure \ref{fig:dist-cats}a.

\subsubsection{Structure generation}
\begin{figure}[!htb]
    \centering
    \includegraphics[width=0.5\textwidth]{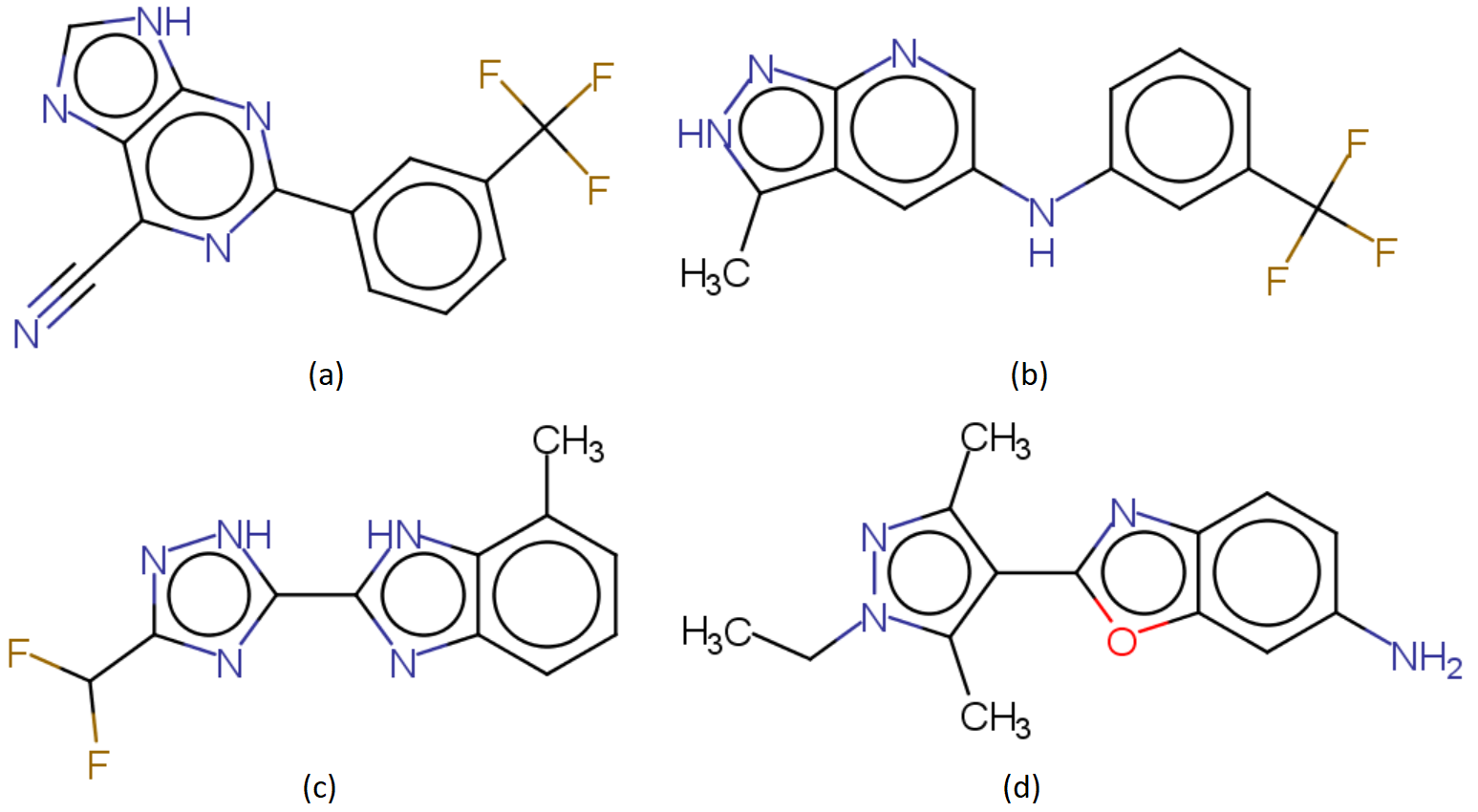}
    \caption{Illustration of similarity between a seed in the CatS set and some generated compounds: (a) The seed; (b) The most similar compound generated from the seed (similarity score=0.45); (c) A compound with a medium similarity score of 0.31; (d) The most different one from the seed (similarity score=0.18). }
    \label{fig:mol-sim-B}
\end{figure}
Using the 2847 compounds in the CatS collected set as seeds and feeding them into the generator network, we generated 1000 distinct compounds for each seed, for a total of 2,847,000 generated compounds. However, there was some duplication among the compounds generated by different seeds, resulting in only 2,080,566 distinct compounds being generated. To determine the novelty of our generated network, the similarity score between each generated compound and its seed is evaluated and depicted in Figure \ref{fig:dist-cats}b.

Similar to the results from the BACE data set, the similarity scores of the generated compounds have a broad range from 0.06 and 1.00. A similarity score of 1.00 was obtained only 12 times in all 2,847,000 generated molecules. In most cases, the similarity scores are very low with an average value of 0.34, indicating there is a lot of diversity among the generated samples. To further verify that the generated compounds are really  different from the seeds, a seed molecule and several generated compounds are shown in Figure \ref{fig:mol-sim-B}. In which, Figure \ref{fig:mol-sim-B}a is one seed, Figure \ref{fig:mol-sim-B}b is the most similar compound generated from the seed with a similarity score of 0.45, Figure \ref{fig:mol-sim-B}c is the compound with a medium similarity score of 0.31, and Figure \ref{fig:mol-sim-B}d is the most different one with a similarity score of 0.18. Obtaining low similarity scores between generated compounds and feeding target is one of the desired features in our GNC model in which the novelty of computer-generated molecules is emphasized.

\subsubsection{Binding affinity screening by 2DFP-DNN}\label{sec:CatS_dist_affinity}

Here, we carry out the 2DFP-DDN model to filter out the ``bad'' generated CatS molecules by the binding affinity criterion. Similar to the BACE compound screening conditions, we use an affinity threshold at -9.56 kcal/mol. Specifically, any molecules with predicted energy higher than that threshold are left out.
As a result, we selected 61,571 potentially ``good'' compounds.

Furthermore, we are interested in the overall distribution of the binding affinity of the generated compounds. Figure \ref{fig:dist-cats}c depicts the distribution of the predicted affinity for all 2,080,566 molecules.
The distribution is fairly close to a Gaussian distribution. Consistent with the similarity score distribution above, the range of their binding affinity prediction is very large, from -4.61 kcal/mol to -12.12  kcal/mol, confirming that  large chemical space is covered. The mean binding affinity is -7.62 kcal/mol. Among the compounds with the smallest predicted binding affinity, 21,283 compounds have binding affinity smaller than -10 kcal/mol, 510 compounds have binding affinity smaller than -11 kcal/mol, and 1 compound has a binding affinity smaller than -12 kcal/mol. These are potentially very highly active compounds.
However, as discussed in Section \ref{sec:BACE_dist_affinity}, there is no free lunch in the development of binding prediction models. 2DFP-DNN predictor is extremely fast in training millions of molecules. However, its accuracy is less competitive in comparison to  3D-based models such as MathDL. Thus, we still utilize the MathDL scoring function to select the most promising drug candidates.

\subsubsection{Clustering analysis of selected compounds}
\begin{figure}[!htb]
    \centering
    \includegraphics[width=0.9\textwidth]{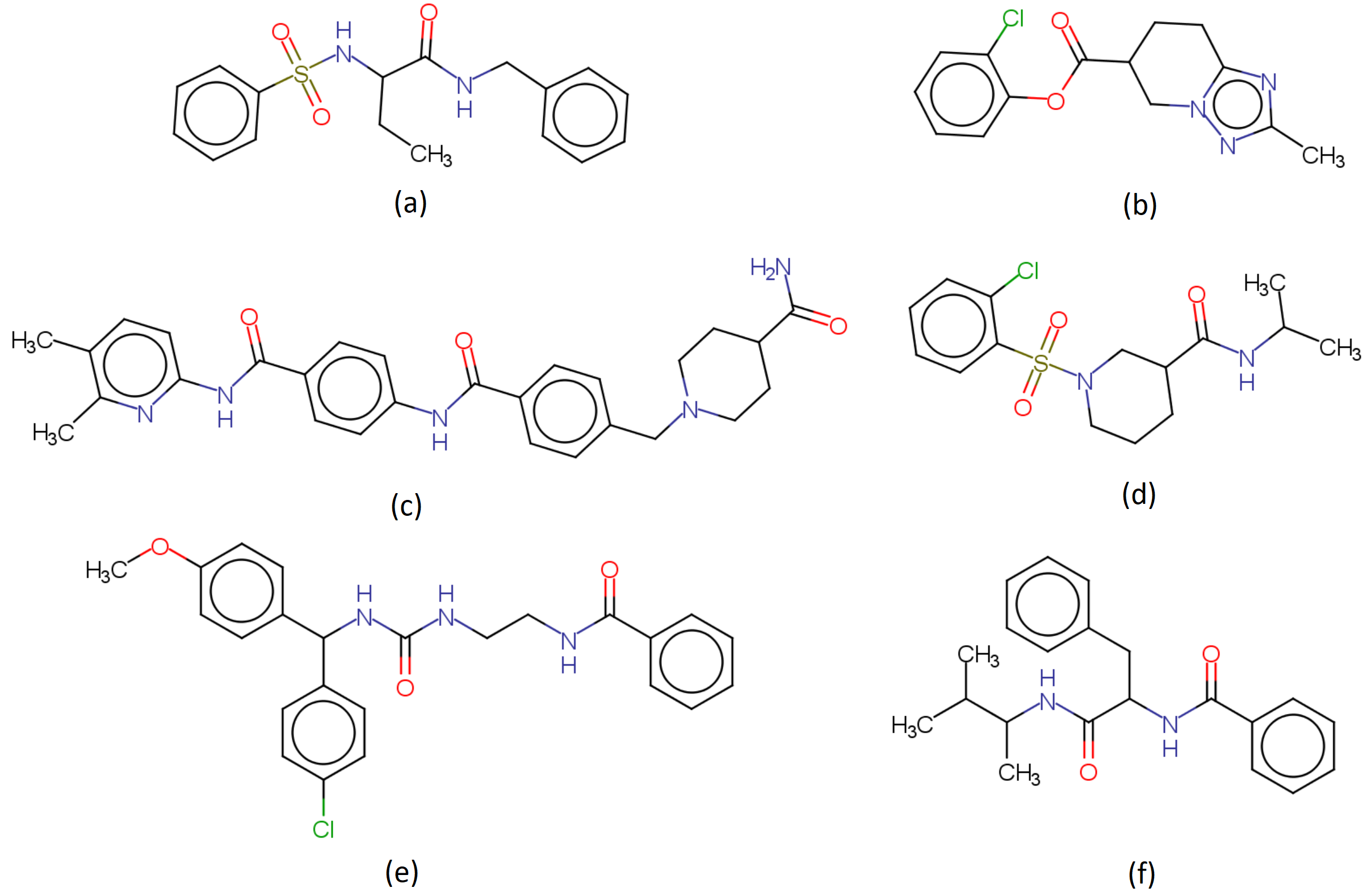}
    \caption{The center of the 6 clusters found in our CatS generated set.}
    \label{fig:clusters-B}
\end{figure}
To illustrate how diverse our generated compounds are, clustering analysis were performed to the 61,571 selected compounds generated by our model. The 6 clusters are found in our generated set, and the center of each cluster is shown in Figure \ref{fig:clusters-B}. The sizes of these 6 clusters are  7077, 13059, 15048, 9221, 6884, and 10282 respectively. Inside these 6 clusters, the average similarity scores to the centers are  0.37, 0.34, 0.34, 0.39, 0.41, and 0.36 respectively, which indicates that there is a significant variation among compounds in each cluster. In addition, the average binding affinity of each cluster  is -10.01, -9.89, -9.91, -9.98, -9.92, and -9.93 kcal/mol respectively. Unlike  the BACE data set, there is not much difference in the average energies between different clusters. Therefore, it is expected to obtain highly potential drug candidates with dissimilar physical and biological chemical properties.

\subsubsection{Binding affinity screening by MathDL}
\begin{figure}[!htb]
    \centering
    \includegraphics[width=0.7\textwidth]{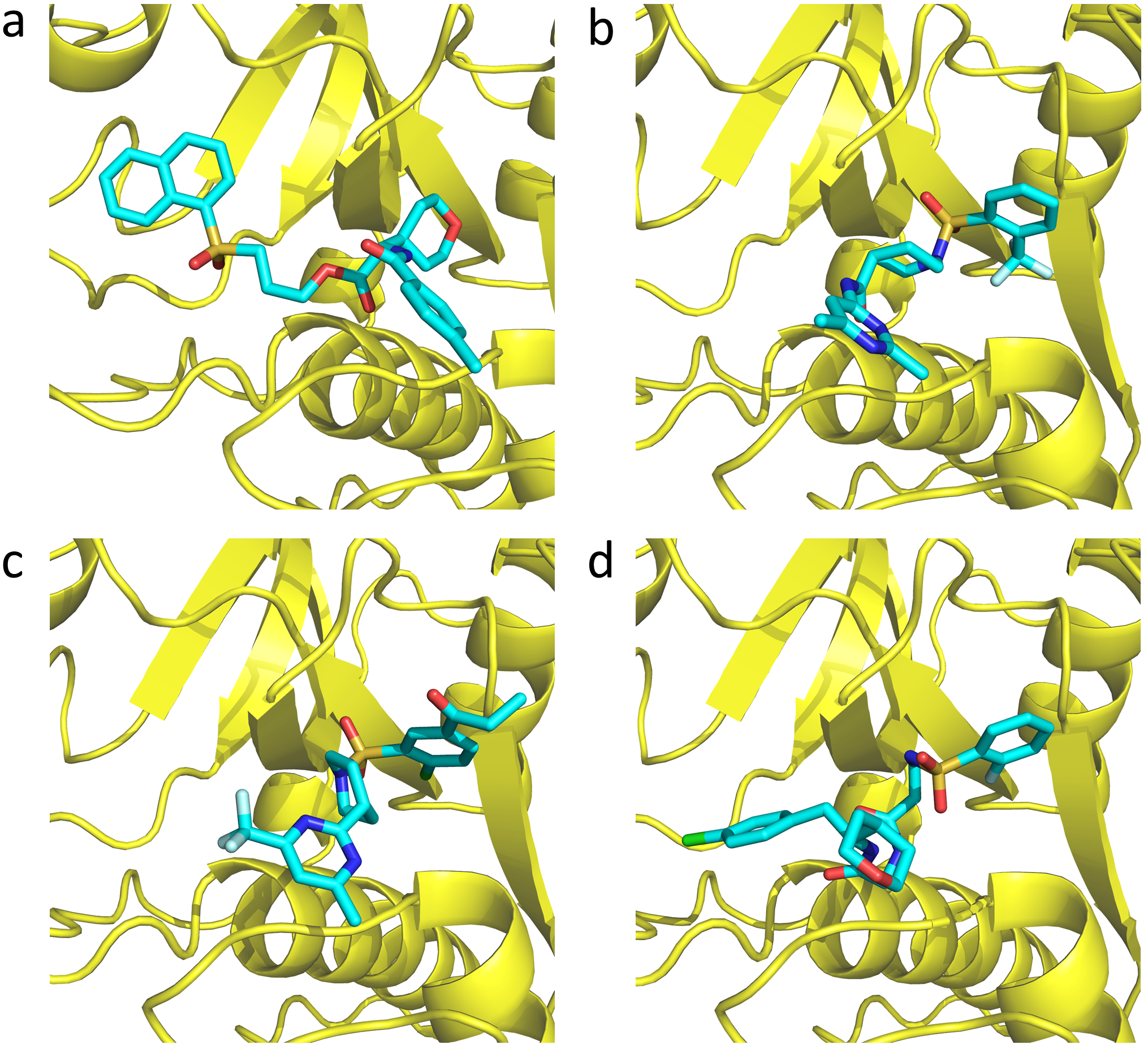}
    \caption{Four generated CatS compounds having the lowest binding affinities predicted by MathDL model. Their 3D structures were predicted by MathPose. Their IDs under our naming system and the predicted energies are, respectively (a) CatS\_gen\_195 (-11.681 kcal/mol); (b) CatS\_gen\_968 (-11.608 kcal/mol); (c) CatS\_gen\_902 (-11.540 kcal/mol); and (d) CatS\_gen\_228 (-11.536 kcal/mol). Their SMILES strings are provided in Table S2, and their corresponding 3D structures are included in File S2.}
    \label{fig:CatS_top4}
 \end{figure}
MathDL here was trained with 2847 seeds used in the generator network. The Pearson's correlation coefficient and Kendall's Tau coefficient on the 10-fold cross-validation (CV) of the training set was found to be $R_p=0.746$ and $\tau=0.577$, respectively. The promising CV performance ensures a well-trained machine learning model. Furthermore, the reliability of the MathDL models on the affinity ranking of the CatS inhibitors has been shown in the Grand Challenges 3 \cite{nguyen2019mathematical} and 4 \cite{nguyen2019D3R} where our models were ranked 1st place among more than 50 teams from over the world.

To further validate the generated molecules, the top 1050 compounds in term of energy indicated by 2DFP-DNN network are re-ranked by the MathDL model. To get the input ready for the structure-based model, the 3D poses of those 1050 molecules are predicted by our MathPose. Under the 2DFP-DNN predictor, the average binding affinity of those 1050 compounds is -11.05 kcal/mol and their affinities range from -10.693 kcal/mol to -12.159 kcal/mol with standard deviation being -0.213 kcal/mol. On the other hand, by utilizing the MathDL model,  the average binding affinity of the selected molecules is -9.27 kcal/mol with a range between -7.008 kcal/m and -11.681 kcal/mol, and the standard deviation is found to be -0.736 kcal/mol. The Pearson's correlation coefficient on the energy prediction for the generated compounds by 2DFP-DNN and MathDL is as low as 0.112 which indicates the disagreement between those two models. That discrepancy was also observed when predicting the affinity ranking of the CatS molecules in the Grand Challenges 4 where the structure-based model MathDL outperformed its ligand-based counterpart. Therefore, MathDL's predicted energies are chosen to select the promising drug candidates among the computer-generated compounds.

The 3D structures of top 4 compounds in term of affinity, namely CatS\_gen\_195, CatS\_gen\_968, CatS\_gen\_902, and CatS\_gen\_228, are plotted in Figure \ref{fig:CatS_top4}. Their reported affinities are, respectively, -11.681 kcal/mol, -11.608 kcal/mol, -11.540 kcal/mol, and -11.536 kcal/mol. Despite   similarly predicted affinities, their structures are quite dissimilar from each other. Specifically, the highest similarity score is 0.297 obtained between CatS\_gen\_968 and CatS\_gen\_902 molecules. While the lowest similarity score is  0.11 evaluated between  CatS\_gen\_902 and CatS\_gen\_228. The statistical information again confirms the ability of our proposed GNC to cover  large chemical space.

\section{Discussions}
Since the chemical space is huge, there is a need to generate a wide variety of novel compounds for all kinds of properties. This work introduces the  GNC  to generate novel molecules, predict their druggable properties, and finally pick up the drug candidates that fulfill the threshold for drug properties such as binding affinity. We discuss a number of issues concerning generative networks.
{
\subsection{Latent space design of new compounds}

 Latent space information can be effectively modified by a variety of methods. In the current work, we propose three approaches,
 including 1) randomized output, 2)  controlled output, and 3) optimized output.  The first approach can certainly create new molecules. We note that some of the new latent space configurations cannot be interpreted by the decoder.

The second approach is designed to discriminate potentially drug-like molecules from potentially inactive ones. Currently, we found that machine learning models built from latent space representations are highly accurate. Therefore, the proposed approach is potentially very useful. Nonetheless, the performance of this method depends crucially on the quality of training datasets.   Additionally, for this approach to effectively control the druggability of the generated compounds, the decoder must be intensively trained with tens of millions of molecules and have a near-perfect reconstruction rate. Achieving a high reconstruction rate for a diverse class of test compounds is a challenging issue in the design of molecular autoencoders.
This issue is under our consideration.

The third approach is introduced to create new compounds with desirable druggable properties.  Similarly to the last method, the success of this approach depends on the quality of training datasets and machine models and the reconstruction rate of the decoder.  Additionally,  reference selection for each drug property is another important issue. It depends on our current understanding and criteria of drug-like molecules.  However, this approach is very promising and will be an important direction for future studies.

Finally, it is noted that the third approach does not depend on the seed configuration. Therefore, its initial latent space distribution can be chosen randomly. As such, this method can be very fast and efficient.
}

\subsection{Generator efficiency}

One challenge traditional pharmaceutical industry faces is that designing new drug candidates is very time-consuming. This low efficiency obviously can not tackle a variety of health crises human being currently encounters, such as drug-resistant infections and fast mutation of viruses, which requires lots of new drugs in a very short time.

Computers are typically  faster than human beings. Therefore, generating new drugs by computers is a potential solution. Such as in our case, just using one K20 Nvidia CUDA GPU card, our generator network can generate 2.08 million and 2.8 million novel compounds for the CatS and BACE targets in less than one week, such task is far beyond human power. Moreover, such process is fully automatic and even does not need human supervision. So, such automatic generators can provide us a huge drug-candidate database rather than some sporadic ones. What is more, just we already showed in this work, combining this generator with reliable automatic DNN predictors, the bulk of drug candidates can be further screened based on the properties predicted by automatic predictors. This whole automatic workflow should be a promising future of the pharmaceuticals industry.

\subsection{Chemical spaces generated by generators}
Our generated compounds are originated from their seeds, some known ligands binding to CatS and BACE from PDBbind database. Thanks to the magic of the autoencoder including some random source, these generated compounds are truly novel and quite far from their seeds: no matter for CatS and BACE, the average similarity scores to the seeds are just around 0.3. This means the two sources of random works and the generator  creates novel compounds rather than just playing some ``copy" games.

More importantly, these generated compounds spread in huge chemical space, this means our generated compounds cover a large range of chemical properties, so it is more possible to hit potential drug candidates. First, the similarity scores to the seeds have a large range, for BACE it is 0.2 to 0.6, for CatS, it is 0.15 to 0.65. Second, our predicted binding affinities also have a wide range, from -5 kcal/mol to -10 kcal/mol or even to -11 kcal/mol. All in all,  the generator is powerful, originating from seeds and but cover a huge chemical space far away from the seeds.

\subsection{Faith vs novelty}
The random noise regulated latent space we designed here can generate lots of novel compounds far from their seeds, this is due to its design:  random sources are included in the model. However, in other words, this generator is not faithful, since the output is quite different from the input. Such architecture is good for our purpose --- what we want is to create broad new compounds from the seeds rather than faithful ones.

However, in another scenario, faith is highly needed.  Griffiths et al. \cite{griffiths2017constrained}, Jin et al. \cite{jin2018junction},  Kusner et al. \cite{kusner2017grammar} and Dai et al. \cite{dai2018syntax} perform Bayesian optimization in the latent space to obtain compounds with desired properties.
We have also designed controlled latent space and optimized latent space in the present work. In these cases, outputs should faithfully reflect the latent space. Otherwise,  optimization in the latent space could not be faithfully passed to the output.  The reconstructing accuracy is a very critical evaluation.  Much effort has already put to reinforce  reconstructing accuracies, such as grammar VAE \cite{kusner2017grammar},  syntax-directed VAE \cite{dai2018syntax} and junction tree VAE \cite{jin2018junction}, to achieve a reconstructing accuracy as high as 0.76. In comparison, the VAE we applied only has a reconstructing accuracy of 0.20.  It means that our outputs are always new compounds. However, our new gated recurrent unit (GRU)-based  autoencoder can achieve a 99\% reconstructing accuracy, which enables us to carry out desirable design in the latent space. The detail of this work will be published elsewhere.

\subsection{Chemical spaces of predicted high binding affinity compounds}
Using our predictors, high binding affinity compounds can be screened. So one concern is whether these high binding affinity compounds spread in a large chemical space or they are similar to each other and in a small range. According to our results, even the numbers of high binding affinity compounds are only {1050 and 99} for CatS and BACE respectively, they are still quite different. In our clustering analysis, these high binding affinity compounds are classified into 6 clusters, the similarities between clusters are only around 0.4.

The large chemical space covered by the high binding affinity compounds is beneficial to drug design. First, a good drug not only depends on binding affinity but also depends on other properties such as toxicity, log P, log S, clearance, etc. A large chemical space means these high binding affinity compounds have different other properties, so there is more chance for them to pass the screenings based on other properties. Additionally,   more types of related drugs are easier to tackle the fast mutation of viruses.

\section{Conclusion}
 In our work, a generative network complex (GNC) is introduced. {We propose three latent-space techniques, including randomized output,  controlled output, and optimized output to generate novel and potential compounds. } Additionally, their physical and chemical properties are predicted by a two-dimensional (2D) fingerprint-based deep learning predictor, and potential drug candidates are preliminarily screened by predicted properties. Moreover, for promising drug candidates, their 3D poses associated with specific protein targets are predicted by our MathPose, one of the most accurate pose prediction schemes according to D3R Grand Challenges, a worldwide competition series in computer-aided drug design \cite{nguyen2019D3R}. Finally, more accurate property estimations based on the 3D poses are performed by our MathDL, a advanced mathematics-based deep learning network, leading to new drug candidates with the desirable drug properties. This automated platform has been used to generate 2.08 million new drug candidates for Cathepsin S and  2.8 million novel compounds for BACE.  For 1050 potential drug candidates for CatS and 99 potential drug candidates for BACE, 3D poses associated with their target proteins have been created to further evaluate their druggable properties. Our framework is designed to create new drugs in silico, so as to save time and reduce cost in drug discovery. {
Designing  gated recurrent unit (GRU)-based autoencoders with near perfect reconstruction accuracies is under our consideration to achieve robust latent space drug design.  }

\vspace{1cm}
\section*{Supplementary materials}
Supplementary materials are available online for potential drug candidates for Cathepsin S  and BACE targets.
\paragraph{TableS1.csv} A list of SMILES strings and predicted binding affinities of 99 potentially active compounds for the BACE target.
\paragraph{TableS2.csv} A lsit of SMILES strings and predicted binding affinities of 1050 potentially active compounds for the  CatS target.

Supplementary materials are available upon request for potential drug candidates for Cathepsin S  and BACE targets (near 70 gigabytes in size).

\paragraph{FileS1.zip} Zip file of 3D structure information of 99 selectively generated BACE compounds and their receptors.
\paragraph{FileS2.zip} Zip file of 3D structure information of 1050 selectively generated CatS compounds and their receptors.

\vspace{1cm}
 \section*{Acknowledgments}
This work was supported in part by  NSF Grants DMS-1721024,  DMS-1761320, and IIS1900473 and NIH grant  GM126189. DDN and GWW are also funded by Bristol-Myers Squibb and Pfizer.

\vspace{1cm}
\section*{References}
\renewcommand\refname{}


\end{document}